# Lattice Expansion

# in Seamless Bi-layer Graphene Constrictions

# at High Bias

*Felix Börrnert,*[*,†,1] *Amelia Barreiro,*[*,‡,2] *Daniel Wolf,*[3] *Mikhail I. Katsnelson,*[4] *Bernd Büchner,*[1] *Lieven M. K. Vandersypen,*[2] *and Mark H. Rümmeli*[§,1,3]

[1]IFW Dresden, PF 270116, 01171 Dresden, Germany, [2]Kavli Institute of Nanoscience, Delft University of Technology, Lorentzweg 1, 2628 CJ Delft, The Netherlands, [3]Technische Universität Dresden, 01062 Dresden, Germany, [4]Radboud University Nijmegen, Heyendaalseweg 135, 6525 AJ Nijmegen, The Netherlands

[†]f.boerrnert@ifw-dresden.de, [‡]ab3690@columbia.edu, [§]m.ruemmeli@ifw-dresden.de

**RECEIVED DATE (to be automatically inserted after your manuscript is accepted if required according to the journal that you are submitting your paper to)**

CORRESPONDING AUTHOR FOOTNOTE [*]These authors contributed equally to this work. [†]f.boerrnert@ifw-dresden.de [‡]ab3690@columbia.edu, Present address: Dept. of Physics, Columbia University, New York, New York 10027, USA, [§]m.ruemmeli@ifw-dresden.de

ABSTRACT  Our understanding of *sp$^2$* carbon nanostructures is still emerging and is important for the development of high performance all carbon devices. For example, in terms of the structural behavior of graphene or bi-layer graphene at high bias, little to nothing is known. To this end we investigated bi-layer graphene constrictions with closed edges (seamless) at high bias using *in situ* atomic resolution transmission electron microscopy. We directly observe a highly localized anomalously large lattice expansion inside the constriction. Both the current density and lattice expansion increase as the bi-layer graphene constriction narrows. As the constriction width decreases below 10 nm, shortly before failure, the current density rises to $4 \cdot 10^9$ A cm$^{-2}$ and the constriction exhibits a lattice expansion with a uniaxial component showing an expansion approaching 5 % and an isotropic component showing an expansion exceeding 1 %. The origin of the lattice expansion is hard to fully ascribe to thermal expansion. Impact ionization is a process in which charge carriers transfer from bonding states to antibonding states thus weakening bonds. The altered character of C-C bonds by impact ionization could explain the anomalously large lattice expansion we observe in seamless bi-layer graphene constrictions. Moreover, impact ionization might also contribute to the observed anisotropy in the lattice expansion, although strain is probably the predominant factor.

MANUSCRIPT TEXT Graphene stands as a unique material for high performance electronic applications.[1] Both its mono-layer and bi-layer form are very attractive for electronic applications. Bi-layer graphene can have similar or distinct properties as compared to its monolayer counterpart depending on the angle of rotation between the two layers.[2] Morozov and coworkers demonstrated that the intrinsic mobility of bi-layer graphene is comparable to that of single layer graphene.[3] However, unlike single layer graphene, an energy gap can be opened in bi-layer graphene in a controlled manner when applying an external electrical field.[4–6] Despite the promise afforded by graphene and bi-layer graphene as building blocks for electronic devices and circuitry, their actual development to date is limited. In part, this is because the technology to achieve this is still lacking due to the need to structure graphene with high (atomic) precision in a reproducible and controlled manner.[7] Graphene and bi-layer graphene are able to sustain remarkably high current densities,[8–11] however when structured as long ribbons, lower current densities as compared to large area graphene are obtained due to electron scattering at edges and reduced thermal conductivity.[12] The production of graphene ribbon based devices with ballistic transport is attractive as it will enable faster devices as well as superior current density limits. However, our understanding of *sp²* carbon based nanostructures at high bias is still emerging. In particular, in terms of the structural behavior of graphene or other *sp²* carbon nanostructures at high bias little to nothing is known.

To this end, we present an *in situ* atomic resolution transmission electron microscopy (TEM) study of a bi-layer graphene nano-constriction at high bias. We observe highly localized lattice expansion inside the constriction that exhibits a uniaxial component of about 5 % and an isotropic component of more than 1 % when the constriction width decreases below 10 nm and the current density rises to $4 \cdot 10^9$ A cm$^{-2}$. The origin of the lattice expansion is discussed.

Details of the device fabrication can be found in the Supporting Information. The chip with the contacted graphene sample is mounted on a custom-built sample holder for TEM with electrical terminals. For imaging, a FEI Titan[3] 80–300 transmission electron microscope with a CEOS third-order spherical aberration corrector for the objective lens is used. It operates at an acceleration voltage of 80 kV to reduce knock-on damage. The image acquisition time was 0.5 s. The electrical characterization was performed using a HP 4140B pA meter / DC voltage source. The micrographs are evaluated using the Gatan DigitalMicrograph software with the Triebenberg package.

In Figure 1 a schematic overview of the *in situ* transmission electron microscope setup is presented in which a sheet of mechanically cleaved few-layer graphene is suspended across two free-standing gold electrodes. The use of TEM allows high degrees of structural information about the constriction to be obtained with relatively high temporal resolution while simultaneously gathering electrical data. The electrical setup provides a voltage bias across the electrodes and monitors/records the current. Upon applying a sufficiently large bias the graphene begins to crack. The cracking process occurs at the centre of the graphene sheet between the electrodes where the temperature rise is greatest. Usually the cracks initiate from the outside edges more or less simultaneously and propagate towards the centre through current induced sublimation of carbon atoms from the crack edges. As the cracks approach the center, a narrow constriction forms. Greater details of the cracking process forming the constriction are available elsewhere[11] and so we do not examine this further. Instead, in this study, we focus on the constriction just before failure while under high current densities. The TEM micrograph provided in figure 1 shows a flake after its width has eroded down from approximately 400 nm to a constriction approximately 2 nm long and 10 nm wide. The edges of the ribbon and crack region are for the most part atomically smooth over a long range and exhibit strong contrast. These features indicate the constriction is bi-layer with closed edges (seamless) and is in keeping with previous observations that current induced cracking leads to bi-layer graphene with closed edges.[11,13] Closed edges add stability by reducing dangling bonds[14] and also add mechanical strength.[15] The reflexes observed from the Fourier transform (FT) of the image confirm crystalline graphene and highlight a rotational stacking fault of around 3° between the two layers. This rotation explains the observed Moiré pattern,[16] as well as the periodic contrast bands that are present across the ribbon (see figure S1 in the Supporting Information). The stacking rotation between the two graphene layers can alter the electronic properties with respect to Bernal (AB) stacked bi-layer graphene. For example, stacking rotations between layers of more than 1.5° are predicted to decouple the two layers electronically inducing a transition from a parabolic to a linear dispersion, characteristic of monolayer graphene.[2] Other studies predict a semimetallic behavior with a small indirect overlap of the valence and conduction band for shifted bi-layer graphene which does not conform to AB or AA stacking.[17] It is also worth noting that the seamless bi-layer constriction is clean, free of unwanted amorphous species and is also highly crystalline which is typical for the current annealing process.[18,19]

We now turn to examining the graphene constriction's lattice whilst at a fixed bias of 2.53 V. As previously stated the current annealing process erodes the constriction with time (e.g. see panels 1 to 5 in figure 2). In this case the constriction width is reduced from 11.3 nm to 7.8 nm. The lower graph in figure 2 shows how the current drops as the constriction narrows down. Since the bias is constant, in essence the resistance is increasing and is an indication that carbon atoms are being sublimed. Knowledge of the constriction's dimensions and current allow us to determine the current density, which changes from 24.5 mA µm$^{-1}$ to 28.5 mA µm$^{-1}$ for the ribbons shown in panel 1 and 5 respectively. This latter value is equivalent to $4 \cdot 10^9$ A cm$^{-2}$, assuming a bi-layer graphene thickness of 0.7 nm. These remarkably high current densities are in line with recent reports.[10,11] Soon after the micrograph aquisition of the constriction shown in panel 5, the nano-ribbon fails and the current drops to zero. The average erosion rate of the constriction's width was 0.077 nm s$^{-1}$ and allows us to extrapolate the ribbon width at failure, which is *ca.* 7 nm and would correspond to a breakdown current density of 30 mA µm$^{-1}$.

The sublimation of carbon atoms from the constriction edges is based on Joule heating. Thus we might anticipate lattice changes, *viz.* thermal expansion. Local crystallographic parameters of the bi-layer ribbon can be accessed from the FT of the image or selected regions of the image. As an example, the inset in panel (b) in figure 3 shows the Fourier transform of the region in the micrograph indicated by the black square. The distance of the spots from the center, i. e. the spatial frequency, represent the directionally resolved lattice parameter (see figure S2 in the supporting information for greater detail). To determine the error, we determine the variation from a series of bias free room temperature reference images that were taken with exactly the same imaging parameters used here. The error is about 1 % (see figures S2 and S3 in the Supporting Information). We do this for each of the micrographs collected over the time frame shown in figure 1. Moreover, we independently examine the constriction and the region just outside by selecting a specific region to obtain the FT as shown in panels (a) and (b) in figure 3. In the ideal case the Fourier transform will produce six spots in an isotropic hexagonal configuration due to graphene's 3 fold symmetry. However, in our case the reflex spots are anisotropic. This anisotropy could be attributed to two reasons. In the first, a distortion of the image due to the aberration correction element may occur.[20] The corrector consists of two multipoles in which the first massively distorts the image and the second then reverses the distortion of the first multipole. This process is known to sometimes leave anisotropic image distortions even after an alignment procedure resulting in an optimized phase plate. Factory alignments aim to keep the distortion below 1%, however, additional distortion can be introduced unintentionally by the user even with an optimized phase plate. However, within a working session the distortion will remain constant so long as the system is not subjected to a further optimization process. In these studies no re-alignment to the $C_s$ correction is applied within a working session so any distortion present can be considered stable.[20] The second distortion process is strain, e.g. due to fabrication.[21] Generally all free-standing graphene membranes are subjected to a degree of strain and we also expect some strain in our graphene membrane to be the most probable scenario. Because of the anisotropy in the spatial frequency of the reflex spots we separate out each of the three directions and plot the (directional) spatial frequency against current density. This is done for two regions; inside the constriction (figure 3, left side) and outside the constriction (figure 3, right side). For each of the series for each direction a least squares fit is applied to highlight the trend in spatial frequency with respect to the current density (or reduction in ribbon width). In all cases the spatial frequency is reduced as the current density increases and this corresponds to a relative lattice expansion. Within the constriction (figure 3, left side) the spatial frequency reduction is noticeably larger in the 11 o'clock direction as compared to the other directions (1 o'clock and 3 o'clock). The 11 o'clock direction is parallel to the direction of the graphene sheet suspended across the gold electrodes suggesting a tensile strain exists across the electrodes. If the change was solely due to tensile strain, contraction would be observed in the other two directions. However, we observe an additional expansion in all directions and this is concomitant with thermal expansion. Outside the constriction where the current density is much smaller, we expect a cooler region. Indeed, the smaller change in spatial frequency with current density indicates a cooler region. Moreover, the changes outside the constriction are approximately the same in all 3 directions. This can be expected since inside the constriction the tension between the electrodes is concentrated in a small area. Similar investigations of the ribbon at an earlier stage where the ribbon width is larger, between 19 nm and 22 nm, show reduced and more parallel changes in the spatial frequency in keeping with a reduced strain and reduced temperature (see figure S4 in the supporting information). These trends confirm that both strain and temperature changes are involved in the relative lattice changes we observe in the seamless bi-layer ribbon. In addition, the spatial relaxation of the two graphene flakes remaining on the electrodes after failure (ribbon rupture), further confirm strain was present (See figure S5 in the supporting information). Within the constriction (figure 3, left side) the changes in spatial frequency correspond to a relative lattice expansion of around 1 to 1.5 % in the 1 o'clock and 3 o'clock directions and a relative expansion of *ca.* 4–5 % in the 11 o'clock direction.

Extrapolating our in-plane lattice expansion of 1 to 1.5 % with temperature for bi-layer graphene from theoretical predictions based on semi-empirical interatomic interaction potentials,[22] yields a temperature between ~ 4000 K and ~ 5500 K. Estimates based on quasiharmonic approximations for graphite show temperatures of more than 7500 K while graphene never expands.[23] One can anticipate the temperature estimate for bi-layer graphene to lie somewhere in between that for graphene and graphite. Calculations show that graphene melts at < 4900 K.[24] Since the constriction is highly crystalline (no melting is observed), this reduces our estimation of the temperature window to between 4000 and 4900 K. Assuming a temperature of 4000 K,[22] one can extrapolate a thermal conductivity, $\kappa$, of *ca.* 190 W (m K)$^{-1}$ (see supporting information). This value does not include any changes introduced by strain, however calculations for single layer graphene with a 5 % strain show only a 30 % increase in $\kappa$.[25,26] This suggests the change in $\kappa$ for our bi-layer constriction due to strain will be rather limited. In essence, our obtained thermal conductivity is an order of magnitude smaller than the published value of 2800 W (m K)$^{-1}$ for bi-layer graphene[27] and other $sp^2$ carbon based materials[28] (see figure S6 in the Supporting Information). This implies that the temperature estimate is too high. On the other hand, if we assume a thermal conductivity of 2800 W (m K)$^{-1}$, as reported for bi-layer graphene[27], a temperature estimate yields a value approximately 630 K. This is significantly below a previously reported temperature of 2000 K for graphene at high bias.[29] In essence; the anomalously large lattice expansion is quantitatively too large to be explained in terms of Joule heating since state-the-art calculations yield values that are too small by an order of magnitude. Thus, the expansion must be due to non-equilibrium effects in the electronic structure. Given the high density of electrons traversing the constriction, a possible non-equilibrium mechanism is impact ionization. Impact ionization is a process in which incoming electrons interact and excite valence electrons into anti-bonding states. Large expansions have been observed by X-ray diffraction and LEED in doped graphite (1 % expansion) and "monolayer graphite" (3 % expansion) resulting from charge transfer from bonding states to anti-bonding states, weakening the C–C bonds.[30,31] Exciton formation in which electrons transfer from bonding states to anti-bonding states may result in lattice expansion.[32] The production of excitons in graphene through impact ionization is reported to be an efficient process in both graphene[33,34] and multi-layer graphene.[35] In addition, high-energy electrons and elevated temperatures are argued to enhance impact ionization rates in graphene.[36] Moreover, any expansion that might arise from impact ionization may be anisotropic. In Ge and GaAs the impact ionization coefficients have been shown to depend on crystal orientation.[37,38] Thus, we postulate preferential excitation (weakening) of C–C bond electrons parallel to the current (so as to conserve momentum) in our constriction. This, in turn, leads to anisotropic lattice expansion, $viz$, impact ionization may contribute to the uniaxial lattice expansion we observe. Thus, we argue impact ionization could play a role in both the observed anomalously large lattice expansion as well as its anisotropy.

In summary, we employ atomic resolution $in$-$situ$ TEM to investigate the structural behavior of graphene bi-layer constrictions at high bias. The studies show a localized lattice expansion in the constriction which increases as the constriction width decreases. For a constriction width below 10 nm, *viz*. shortly before failure, a lattice expansion above 1 % is found. The anomalously large lattice expansion cannot be fully explained by thermal expansion. Impact ionization weakening the C–C bonds could explain the observed expansion. Moreover, the lattice expansion is anisotropic which can be attributed to strain, however, in addition, impact ionization may also contribute to the observed uniaxial lattice expansion since impact ionization coefficients depend on crystal orientation.

ACKNOWLEDGMENT We thank S. M. Avdoshenko, M. Lazzeri, and H. Sevinçli for helpful discussions. F. B. acknowledges the DFG (RU1540/8-1), A. B. the Agència de Gestió d´Ajuts Universitaris i de Recerca de la Generalitat de Catalunya (2010_BP_A_00301). A. B., M. I. K. and L. M. K. V. the Dutch Foundation for Fundamental Research on Matter (FOM) and M. H. R. the EU (ECEMP) and the Freistaat Sachsen.

SUPPORTING INFORMATION

**Supporting Information Available**: Details of the device fabrication and calculation of the thermal conductivity are provided. In addition supporting TEM data on the constriction are available. This material is available free of charge via the Internet at http://pubs.acs.org.

FIGURE CAPTIONS

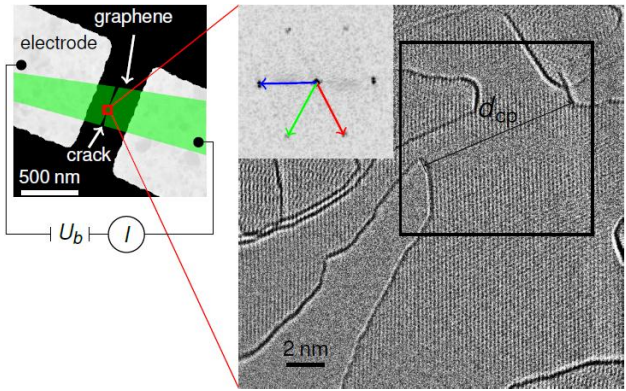

**Figure 1.** Schematic overview of the setup. left — scanning TEM image of the electrodes with a sketch of an overlying graphene ribbon with a central constriction. right — TEM image of the bilayer graphene constriction at high bias. The inset is the FT of the marked region in the micrograph. The three arrows indicate the directions in which intensity profiles were taken, with the color code used in figure 3.

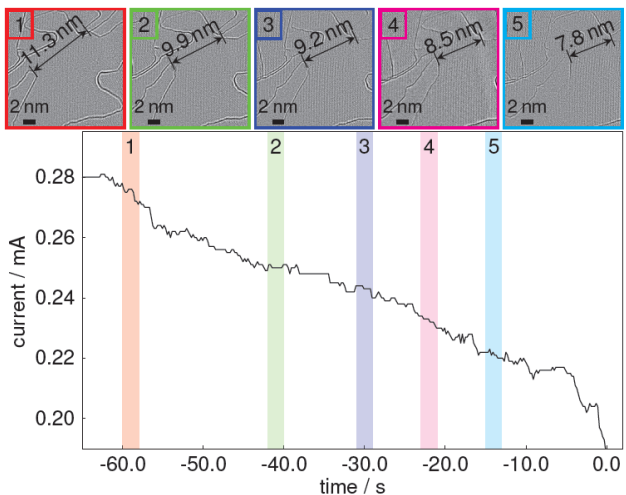

**Figure 2.** Series of micrographs at high bias. The plot shows the measured current through the constriction. The transparent bars indicate when the respective series of color coded micrographs were taken. At time equal to zero the ribbon failed and the current rapidly dropped to zero (not shown in the plot).

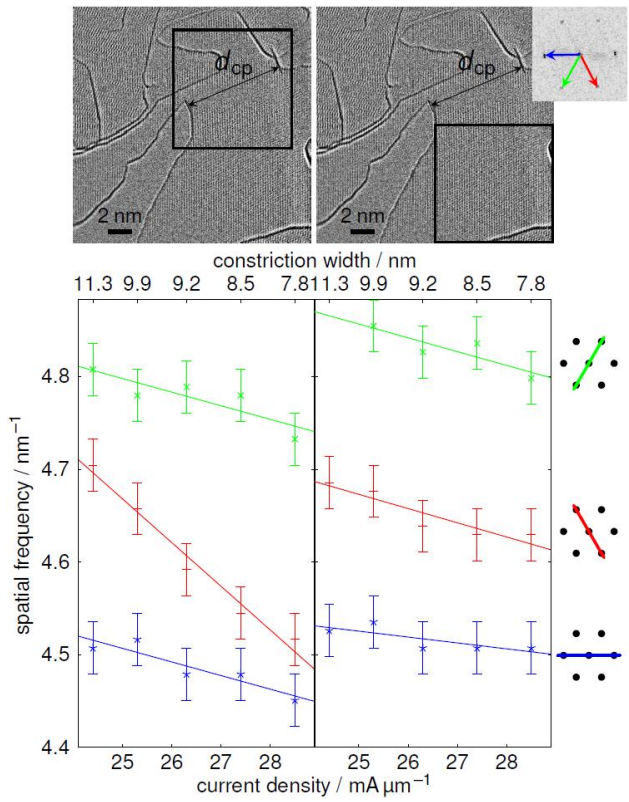

**Figure 3.** The plots show the measured spatial frequencies of each direction separately measured in the FT of the respective TEM micrographs (top) directly in the constriction area (left) and in the lower right corner outside the constriction (right). The solid lines are guides for the eye.

SYNOPSIS TOC

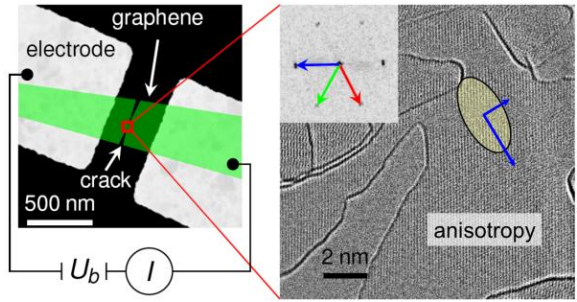

# Lattice Expansion

# in Seamless Bi-layer Graphene Constrictions

# at High Bias

## ―Supporting Information―

*Felix Börrnert,*[*,†,1] *Amelia Barreiro,*[*,‡,2] *Daniel Wolf,*[3] *Mikhail I. Katsnelson,*[4] *Bernd Büchner,*[1] *Lieven M. K. Vandersypen,*[2] *and Mark H. Rümmeli*[§,1,3]

[1]IFW Dresden, PF 27 01 16, 01171 Dresden, Germany, [2]Kavli Institute of Nanoscience, Delft University of Technology, Lorentzweg 1, 2628 CJ Delft, The Netherlands, [3]Technische Universität Dresden, 01062 Dresden, Germany, [4]Radboud University Nijmegen, Heyendaalseweg 135, 6525 AJ Nijmegen, The Netherlands

[†]f.boerrnert@ifw-dresden.de, [‡]ab3690@columbia.edu, [§]m.ruemmeli@ifw-dresden.de

Details of the device fabrication:

Double side polished Si wafers with 300 μm thickness and a 1.3 μm thermal $SiO_2$ on both sides are used. An etch mask consisting of 4.2 μm SiO2 and 540 nm Si is grown on the backside of the wafer by plasma-enhanced CVD. A thick ZEP resist mask is used for patterning 50 × 50 μm² squares which are etched through the entire Si layer by means of deep reactive ion etching[1] leaving free standing $SiO_2$ membranes on the front side of the wafer. Electrodes are patterned on the $SiO_2$ membranes using electron beam lithography and subsequent evaporation of Cr/Au (10 nm / 150 nm). The $SiO_2$ membranes are removed by HF etching.

Graphene flakes are obtained by mechanical exfoliation of kish graphite on Si wafers coated with a 280 nm oxide layer. PMMA is spun on the wafers with graphene flakes on top. The chips are then placed in a NaOH solution and remain floating on top of the solution until the NaOH intercalates between the PMMA film and the wafer. Thereby the PMMA film with the graphene flakes attached to it is released from the wafer. The PMMA film floats on top of the solution while the wafer sinks to the bottom of the beaker.[2] Then the PMMA film is then transferred to a Petri dish with water and a selected graphene flake is aligned onto the electrodes while continuously decreasing the water level.[3] The regions of the PMMA film where the graphene and the Au electrodes overlap are strongly overexposed by means of electron beam lithography so that the PMMA crosslinks, thus clamping the graphene flake to the electrodes. Subsequently, the PMMA is dissolved in acetone and boiling IPA.

The final device is then glued onto a chip carrier with conducting carbon tape and wire bonded to the chip carrier. The chip carrier is then mounted onto a custom built TEM holder that is electrically contacted to the chip carrier.

Comments on calculating the thermal conductivity

To estimate the thermal conductivity at a given temperature inside the constriction or *vice versa* we use a two-dimensional heat dissipation approach.

Within the estimate we assume all of the power applied to the device is released in the constriction which yields a maximum input power of 0.6 mW. The losses in the constriction can be diffusive or radiative, however we neglect radiative losses since the Stefan-Boltzmann law shows an emission power in the order of $10^{-10}$ W at 2000 K. Thus, the two-dimensional energy dissipation in a material with a thermal conductivity $\kappa$ over a distance of $R_0$ to the environmental temperature $T_0$ is given by:[4]

$$D = \frac{4\pi \kappa t (T - T_0)}{0.58 + 2\ln\left(\frac{2R_0}{d}\right)} \quad (1)$$

where $t$ is the sheet thickness, $T$ the temperature inside the constriction, and $d$ the diameter of the hot spot. A hot spot with a 10 nm diameter is reasonable and is used for $d$. The distance to the gold electrodes is ~500 nm to each side which we use as our value for $R_0$, and for the bi-layer graphene thickness we apply a value of 0.7 nm.

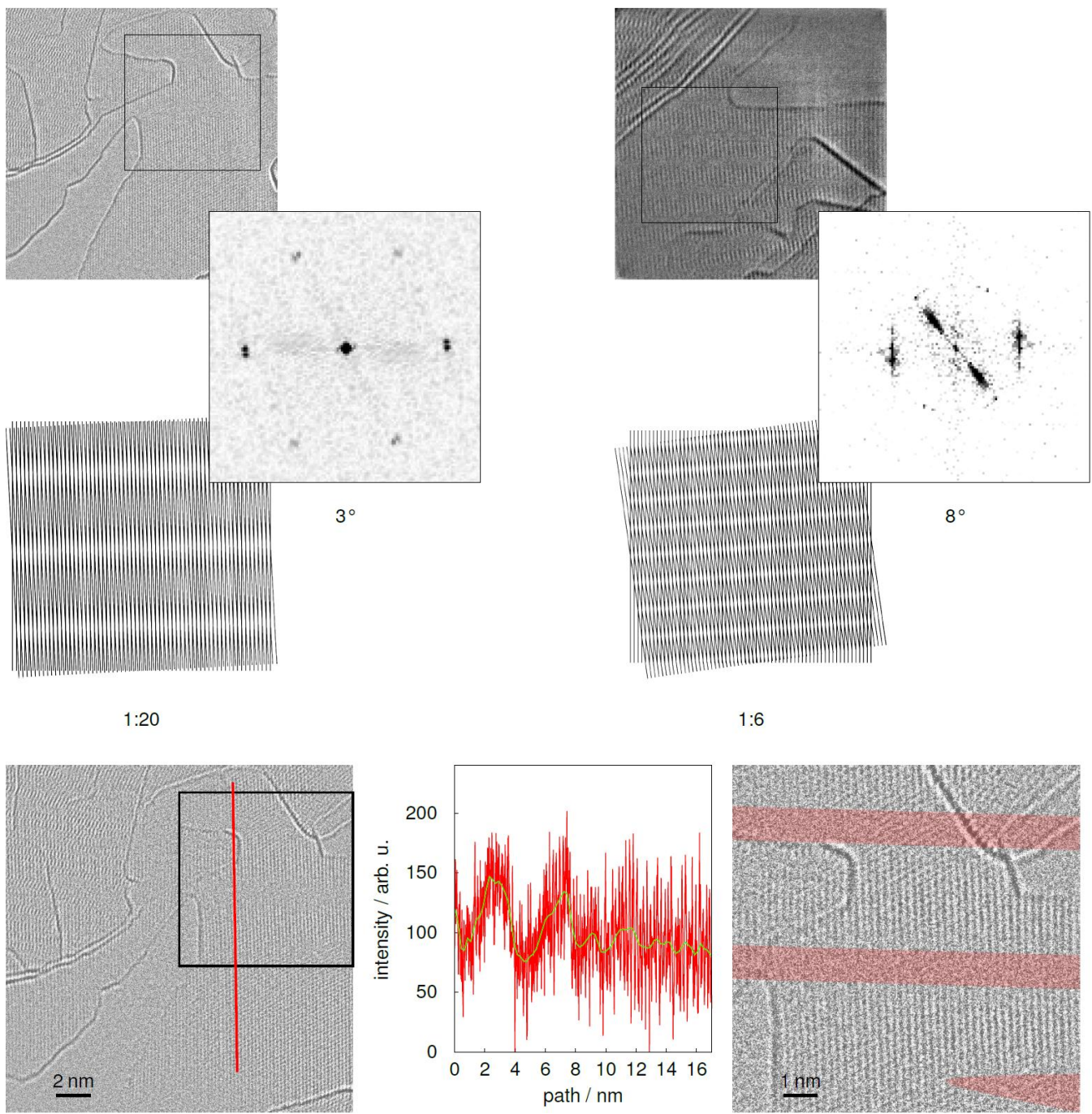

**Figure S1.** Evidence for bi-layer graphene: First, a strong contrast at the edges is observed. Second, on the top right side is a frame earlier in the series where the constriction is still too large for the field of view. The signal of rotated bi-layer (probably from a rotated grain) is clearly seen in the FT. Below, there is a crudely simulated Moiré pattern matching the ratio of lattice periodicity vs. contrast modulation periodicity. On the left hand side a simulated Moiré pattern with the original constriction image, using a rotation of 3°, is approximated from the respective FT. At the bottom a line profile in the intensity valley of the horizontal periodic lattice signal is displayed, showing more intensity in the low-contrast regions. On the right, the magnified section of the micrograph shows the regions of low contrast.

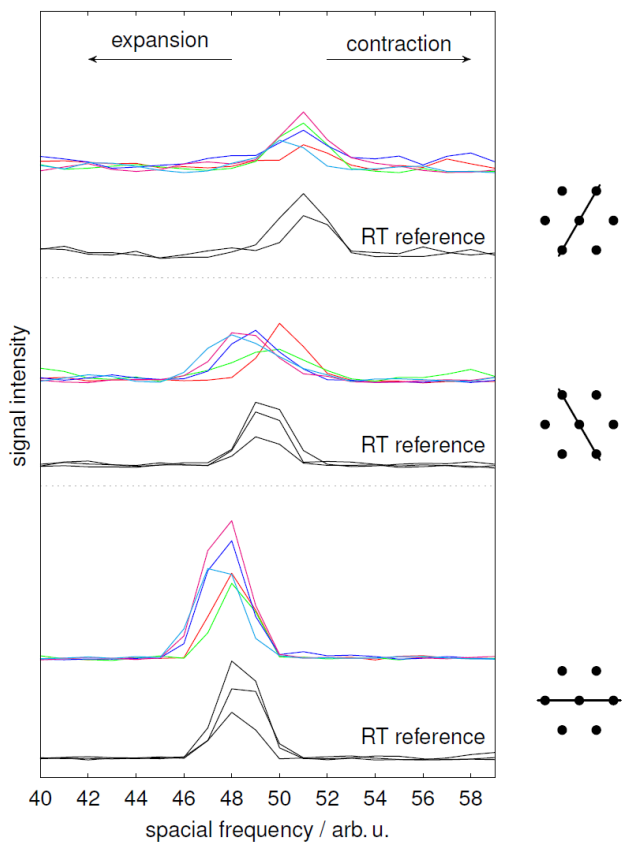

**Figure S2.** The original intensity profiles in the Fourier transform (FT) the data of figure 3 were taken from. The three vertically offset panels show the intensity profiles in the FT for each direction separately as indicated by the pictograms on the right. The color code is the same as in figure 2, i. e. starting from red over green, blue, and magenta to cyan corresponding to a current density from 24.5 mA µm$^{-1}$ to 28.5 mA µm$^{-1}$. In each panel room temperature profiles are shown in black. To determine the error bars used in the spatial frequency data (figure 3 in the main text and figure S4) we determine the variation from a series of bias free room temperature reference images that were taken with exactly the same imaging parameters as used for figures 3 and S4. The error is about 1 % as seen above in figures S2 and S3 (below).

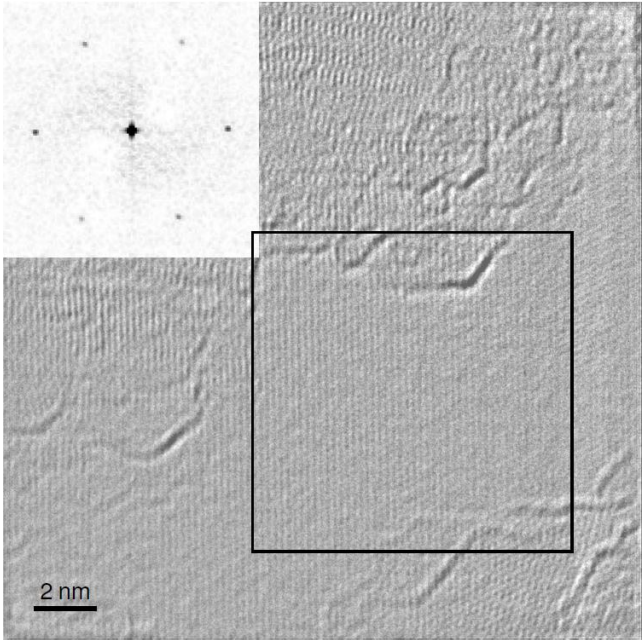

**Figure S3.** Micrograph showing the graphene at zero bias. The inset is the FT of the region marked by a black square in the micrograph.

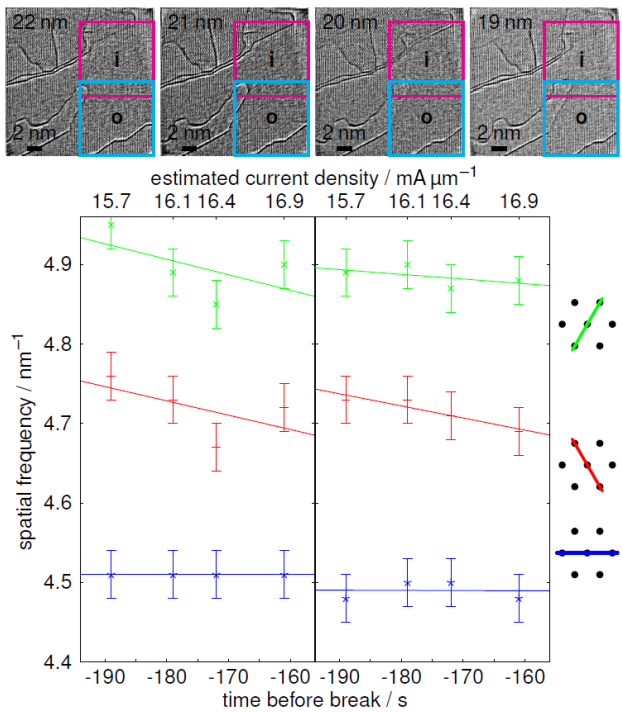

**Figure S4:** Series of TEM images two minutes before the final series of images of the constriction. Top— subsequent micrographs with the estimated width of the constriction indicated at the top. The estimation was done by assuming a constant erosion rate and extrapolating from the final series. The two boxes drawn in each micrograph indicate the regions the signals for the evaluation below stem from. The upper is the "inside" region (left in the plot below) and the lower the "outside" region (right in the plot below). Bottom — spatial frequencies for the respective regions indicated in the series above. The solid lines are guides for the eye. The estimated current density comes from the extrapolated constriction width in conjunction with the measured current at the time when the respective micrograph was taken.

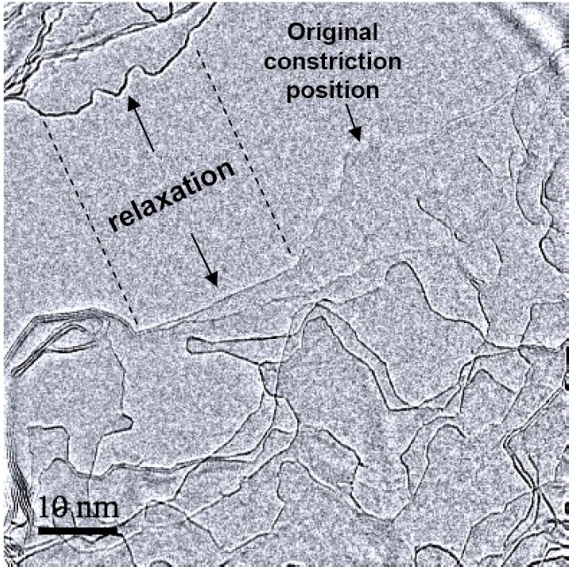

**Figure S5:** TEM image of the constriction region immediately after rupture of the constriction. The relaxation is highlighted as well as the original position of the constriction.

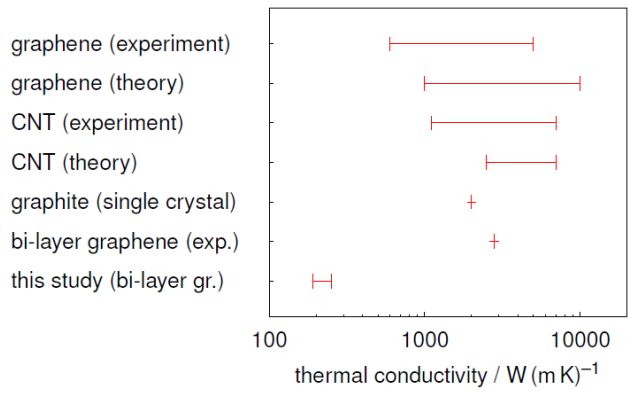

**Figure S6.** Thermal conductivity values for different $sp^2$ carbon species (after ref. 27 in the main text) As compared to the calculated value from our bi-layer graphene constriction at high bias.